\documentclass[conference]{IEEEtran}
\usepackage{cite}
\ifCLASSINFOpdf
  \usepackage[pdftex]{graphicx}
\else
\fi
\usepackage{amsmath}
\usepackage{amssymb}
\usepackage{algorithm}
\usepackage{algorithmic}

\usepackage{array}
\usepackage[caption=false,font=normalsize,labelfont=sf,textfont=sf]{subfig}
\usepackage{dblfloatfix}
\newcommand\blfootnote[1]{
    \begingroup
    \renewcommand\thefootnote{}\footnote{#1}
    \addtocounter{footnote}{-1}
    \endgroup
}
\let\MYorigsubfloat\subfloat
\renewcommand{\subfloat}[2][\relax]{\MYorigsubfloat[]{#2}}
\usepackage{url}
\begin{document}
\title{Using Random Codebooks for Audio Neural AutoEncoders}
\author{Benoît Giniès, Xiaoyu Bie, Olivier Fercoq, Gaël Richard\\
LTCI, T\'el\'ecom Paris, Institut polytechnique de Paris, Palaiseau, France \\
E-mail: \textit{firstname.lastname@telecom-paris.fr}
}
\maketitle
\begin{abstract}
Latent representation learning has been an active field of study for decades in numerous applications. Inspired among others by the tokenization from Natural Language Processing and motivated by the research of a simple data representation, recent works have introduced a quantization step into the feature extraction. In this work, we propose a novel strategy to build the neural discrete representation by means of random codebooks. These codebooks are obtained by randomly sampling a large, predefined fixed codebook. We experimentally show the merits and potential of our approach in a task of audio compression and reconstruction. 
\end{abstract}
\IEEEpeerreviewmaketitle

\blfootnote{This work was funded by the European Union (ERC, HI-Audio,
101052978). Views and opinions expressed are however those of the author(s) only and do not necessarily reflect those of the European Union or the
European Research Council. Neither the European Union nor the granting
authority can be held responsible for them.}

\vspace{-10pt}

\section{Introduction}
\IEEEPARstart{T}{he} extraction of a meaningful and compact representation of the input data is an essential step of modern machine learning based systems. This representation is expected to extract relevant information about the input data to ease the resolution of a target task such as audio classification \cite{Richard-2013-perceptual}, speech enhancement \cite{wang2020selfsupervised-speech} or audio inpainting \cite{Adler-2012-impainting}.
Numerous architectures have been proposed to obtain such a representation for audio signals in a supervised or unsupervised manner \cite{Abdelrahman-review-speech, parekh2019weakly}.
The objective is usually to express the input data under the form of a continuous latent representation which is then further processed by subsequent blocks for the downstream task.
For example, Variational Auto Encoders (VAE) \cite{kingma2014auto,rezende2014stochastic} are very popular models which allow to fit a probabilistic distribution to input data, and randomly sample a corresponding latent representation from it. 

Traditionally, the goal has been to extract a continuous representation from the input data, but there is nowadays a strong trend towards obtaining a discrete representation which has many advantages for data modeling, prediction or generation. For instance, in the VQ-VAE model introduced in \cite{van2017neural}, the latent representation, learned in an unsupervised manner, is quantized using vector quantization exploiting a so-called dictionary (or \textit{codebook}) of tokens (or \textit{codewords}).
Many variations of this model exploiting multiple subdictionnaries were then introduced either in a global hierarchical multiresolution quantization scheme \cite{razavi2019generating,williams2020hierarchical,takida2023hq} or in a successive residual quantizations framework \cite{zeghidour2021soundstream,defossez2022high}.
The latter model is particularly efficient for audio coding and demonstrated excellent performances in sound generation by incorporating an autoregressive language model.

Nevertheless, discrete neural approaches in general face several challenges: first, they are subject to a suboptimal exploitation of the codebooks, the so-called \textit{codebook collapse} problem and they usually rely on ad-hoc heuristics to mitigate this collapse problem; second, there is no clear evidence that all codebooks need to be explicitly learned; and third despite their inherent advantage compared to approaches based on a continuous latent representation, they may still suffer from limited generalization capabilities.

In this work, we propose an alternative strategy to build the discrete latent representation. Inspired by the work of 
Moussallam et al. on Sequential Subdictionaries Matching Pursuits \cite{moussallam2012matching}, 
the core idea of our method is to obtain successive small dictionaries randomly sampled from a large fixed dictionary.  

Using the recent Descript Audio Codec (DAC) model \cite{kumar2024high} as a strong baseline, our preliminary experiments show that this novel strategy for codebook design~:
\begin{itemize}
    \item is robust to the codebook collapse problem, 
    \item has potential to avoid the learning of some of the codebooks, leading to a gain of complexity while maintaining compression efficiency and audio reconstruction quality.
\end{itemize}

The paper is organised as follows: we first describe the related work (section \ref{sec-related}), before describing in detail our approach (section \ref{sec-approach}). We then present our experimental plan in section \ref{sec-experiment} and discuss the results obtained in sections \ref{sec-results} and \ref{sec-discussion}. Finally, we highlight some future work and suggest some conclusions in section \ref{sec-conclusion}.

\section{Related work}
\label{sec-related}

\subsection{VAE for audio neural modeling}
The objective of the VAE~\cite{kingma2014auto,rezende2014stochastic} is to fit a probabilistic distribution to input data, by jointly learning the latent generative modeling and the variational posterior distribution. 
Whereas the original VAEs were primarily applied to image data, subsequent studies have expanded their use to audio data modeling~\cite{blaauw2016modeling} and showcased their effectiveness for various downstream tasks such as speech enhancement~\cite{bie2022unsupervised}, voice conversion~\cite{lian2022robust} and text-to-speech generation~\cite{kim2021conditional}. Meanwhile, the conventional VAE model relies on single latent space  with the \textit{i.i.d} assumption, which motivates researchers to investigate hierarchical~\cite{vahdat2020nvae} and disentangled~\cite{hsu2017unsupervised} latent space, and dynamical modeling~\cite{bie2021benchmark}.

\subsection{Feature quantization and VQ-VAE}
Besides learning continuous representation for audio data, recent studies show the interest of learning discrete, or quantized, representations~\cite{lakhotia2021generative}.
While forcing the model to ignore the irrelevant information during quantization, the discrete representations can then be related to high level concepts easily understandable by humans such as  phonemes in speech or notes in music. Injecting the feature quantization to a VAE leads to the well-known VQ-VAE model~\cite{van2017neural}, that has been used to generate images~\cite{razavi2019generating} and audio~\cite{dhariwal2020jukebox}. More recently, the residual VQ-VAE~\cite{zeghidour2021soundstream,defossez2022high} has been introduced for audio coding and modeling, which also shows impressive performance on audio generation~\cite{borsos2023audiolm}.

\subsection{Codebook collapse}
Despite the promising results in many tasks of generating complex data, the standard training approach frequently encounters codebook collapse, \textit{i.e.} only a portion of available codes are actually utilized, largely restricting  the quality of the discrete latent representations. To mitigate this problem, many techniques have been proposed, such as exponential moving average (EMA) for codebook update~\cite{van2017neural}, codebook reset~\cite{williams2020hierarchical}, a stochastic variant (SQ-VAE)~\cite{takida2022sq}, and factorizing the codes with L2-normalization~\cite{kumar2024high}.

\section{Random Residual Vector Quantization}
\label{sec-approach}

Our main goal in this work is to explore an alternative strategy for defining, using and training the codebooks used in  residual vector quantization (RVQ). As discussed above, dictionary learning is an active area of research, since the training procedures proposed in \cite{van2017neural} and used in \cite{defossez2022high} trigger codebook collapse. Furthermore, given the generative potential of such a quantized representation~\cite{lakhotia2021generative,borsos2023audiolm}, there is a clear interest to build expressive and general purpose dictionaries.

A straightforward modification along these lines would be to increase the size of the quantization codebooks. However, this would greatly increase the computation complexity (at inference and training) and the bitrates (leading to less efficient audio codecs), and would not solve any codebook collapse issue.

We rather propose an alternative strategy inspired by the Sequential Subdictionaries Matching Pursuits algorithm introduced in \cite{moussallam2012matching}.
In this algorithm, a signal is reconstructed by a sum of tokens that are iteratively selected from a sequence of dictionaries. The randomness comes from the fact that, at each step, the best token for reconstruction is selected from a small sub-dictionnary built as a random subsampling of a much larger dictionary. 
Experiments have shown in the context of Matching Pursuit that by proceeding this way, the quality of the reconstruction is almost equivalent to that obtained by using the whole dictionary, for a much lower complexity.

Furthermore, the observations made in \cite{takida2023hq}, as well as the observations we made on DAC model \cite{kumar2024high}, hint that, in a hierarchy of quantizers, deeper quantizers only bring fine grain information to the reconstruction (similar to encoding noise). Thus, it seems relevant to apply the random sampling procedure only to deeper quantizers, as they encode less specific information. In our case, we then obtain these high level quantizers by randomly subsampling a significantly larger codebook, while guaranteeing non-redundant sub-sampling for each quantizer. This permits to directly account for the similar nature of tokens at high level and contribute to enforce generalizability.

As the big codebook is supposed to be large, compared to the initial size of the trained codebooks, we advocate that it does not need to be trained, which is an important advantage for controlling complexity.
To further detail our approach, let's say we aim to quantize $x \in \mathbb{R}^D$ using the codebook $\mathcal{B} = (\beta_i)_{i\leq N} \in (\mathbb{R}^D)^N$. The token is selected following $\arg\min_{\beta \in \mathcal{B}} ||x - \beta||_2^2$. Let's now define $\mathcal{B}_{big} \in (\mathbb{R}^D)^{N_{big}}$ a big codebook, instead of computing $\arg\min_{\beta \in \mathcal{B}_{big}} ||x - \beta||_2^2$, we will randomly extract $\mathcal{B}_s \subset \mathcal{B}_{big}$ of size $s$, and compute $\arg\min_{\beta_s \in \mathcal{B}_{s}} ||x - \beta_s||_2^2$. A schematic description of that process is introduced in Algorithm \ref{model} and in Figure \ref{RandRVQ}.

\begin{algorithm}
\caption{Random RVQ model}\label{model}
\begin{algorithmic}
\STATE \textbf{Input:} Input signal $x \in \mathbb{R}^D$ 
\STATE Trainable codebooks $(\mathcal{B}_i)_{i\leq n_t}$ sampled from $\mathcal{N}(0,\mathcal{I}_D,N_t)$
\STATE Big fixed codebook $\mathcal{B}_{big}$ sampled from $\mathcal{N}(0,\mathcal{I}_D,N_{big})$
\STATE Sampling size $s$
\STATE Number of random quantizers $n_r$
\STATE \textbf{Output:} Quantization tokens $(\Bar{\beta}_i)_{i \leq n_t + n_r}$
\STATE \textbf{Process:}
\STATE $residual \gets x$
\FOR{$i \leq n_t + n_r$}
    \IF{$i \leq n_t$}
        \STATE $\Bar{\beta}_i \gets$ NearestNeighbour($residual$, $\mathcal{B}_i$) 
    \ELSE
        \STATE $\mathcal{B}_s \gets$ UniformRandomSampling($\mathcal{B}_{big}$, $s$)
        \STATE $\Bar{\beta}_i \gets$ NearestNeighbour($residual$, $\mathcal{B}_s$)
    \ENDIF
    \STATE $residual \gets residual - \Bar{\beta}_i$
\ENDFOR
\RETURN $(\Bar{\beta}_i)_{i \leq n_t + n_r}$
\end{algorithmic}
\end{algorithm}

\vspace{-1pt}

\section{Experiments}
\label{sec-experiment} 
\subsubsection{Architecture}
To evaluate our approach, we select the same use case (i.e. \textit{audio reconstruction}) as the EnCodec \cite{defossez2022high} and DAC \cite{kumar2024high} models. The latter architecture is currently the state of the art for audio compression and reconstruction. Just as for EnCodec, the DAC model is formed of an encoder module, a quantization module and a decoder module. The encoder module extracts continuous 2-D latent features from the input 1-D audio data, then the quantization module discretizes the latent representation and feeds the result to the decoder module which builds a reconstruction of the input signal. The encoder and decoder modules are stacks of convolutional and down/up sampling layers, the quantization module is shaped as a residual vector quantizer.

The quantization module is formed of 9 quantizers, which are organized in a pile-like structure, each quantizer computing a new residual after quantization, and feeding it to the next quantizer. 
Our Random RVQ approach is potentially applicable to all quantizers, but in this work we limit our study to the replacement of the 4 to 6 last quantizers, which given their position in the quantization pile, mostly encode noise.

\begin{figure}[!t]
\centering
\includegraphics[width=0.80\columnwidth]{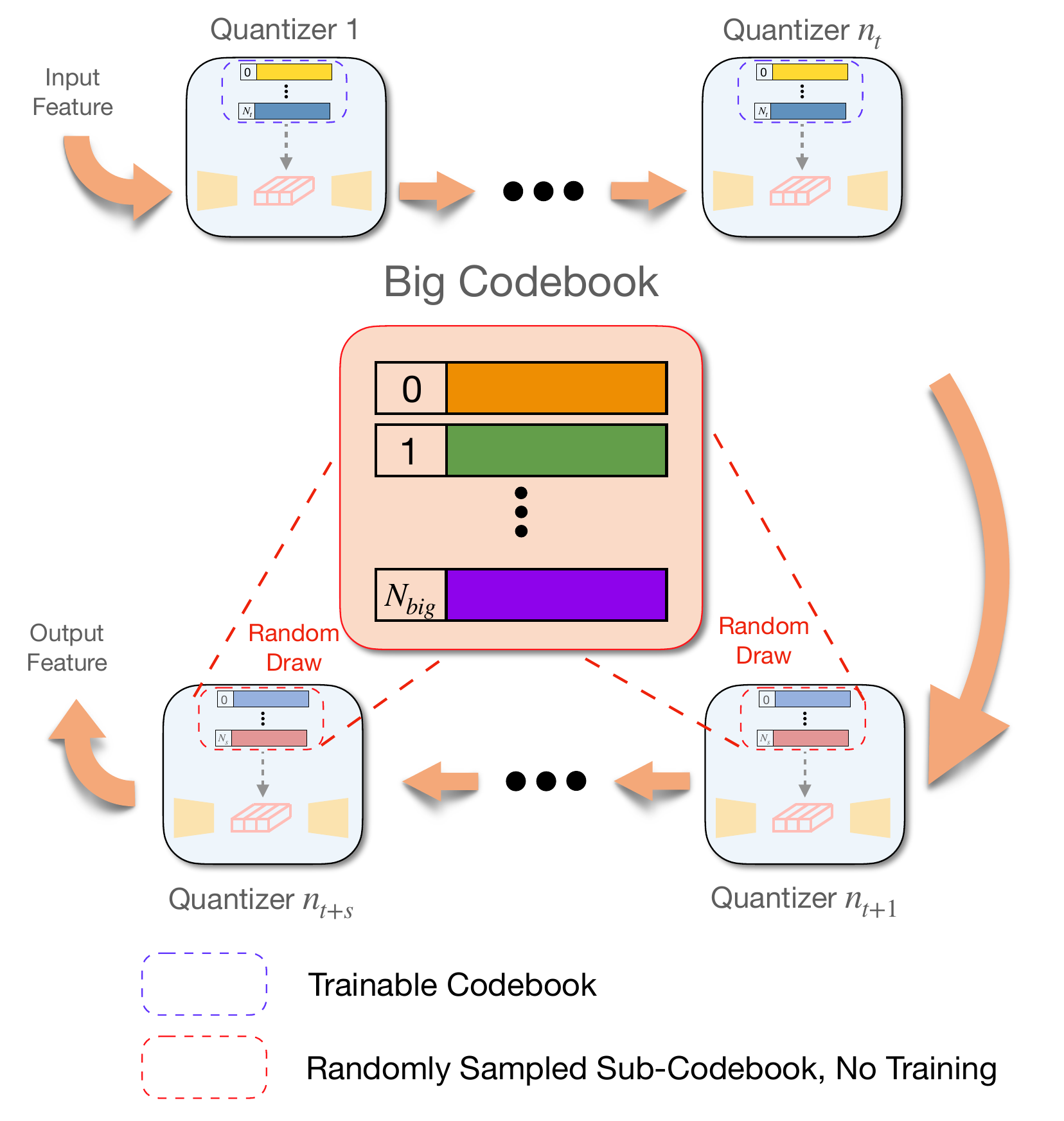}
\caption{The illustration of Random RVQ. Same as DAC~\cite{kumar2024high}, it contains nine layers of quantizer. However, the first five layers are trainable and the last four layers are randomly sampled from a large codebook, while all layers are trainable in DAC. The dimension reduction (orange rectangles in each quantizer) represents the codebook collapse mitigant operations.}
\label{RandRVQ}
\end{figure}

\subsubsection{Data}
The DAC model \cite{kumar2024high} is initially trained on mixed audio data: music, speech and environmental sounds extracted from 8 underlying datasets. In this work, to evaluate the potential of our approach, we trained our model  on only one of these 8 datasets: MUSDB18 \cite{musdb18}. It is composed of a train set of 100 music tracks, and a test set of 50 music tracks, all sampled at 44.1 kHz with a total length around 10 hours. To obtain a meaningful baseline, we retrained the DAC model on that dataset only. It can be noted that only a very limited decrease of performance was observed after re-training which confirms that the DAC model can be considered as a strong baseline.

\vspace{-1pt}

\subsubsection{Evaluation}
We kept the same losses as in DAC~\cite{kumar2024high} for the training of the randomized model: generative and feature matching losses (adversarial), a multi-scale mel loss and codebook and commitment losses. We keep the same balancing between the losses as introduced in DAC~\cite{kumar2024high}.

To evaluate a model, we also applied the same metrics as the one used for the training of DAC: waveform loss, stft loss, mel loss, scale invariant signal-to-distortion ratio (SI-SDR) as introduced in~\cite{le2019sdr}, and the ViSQOL value, a perceptual audio quality assessment introduced in~\cite{chinen2020visqol}.

To monitor the usage of the codebooks, and therefore assess the potential codebook collapse, we report the perplexity of each codebook. 
The perplexity of a codebook $\mathcal{B}$, composed of $N$ tokens $(\beta_i)_{i=1...N}$, is defined as:
$$PP(\mathcal{B}) = e^{-(\sum_{i=1}^N \frac{1}{n_i} \ln(\frac{1}{n_i}))}$$
where $(n_i)_{i=1...N}$ represent the occurences of tokens used for a given test set.
Such a metric converges towards $N$ when the codebook is equally partitioned for quantization. In the case of codebook collapse, the perplexity takes low values.

We also measure the training time, as a statistical estimator for complexity.\\

\vspace{-2pt}

\subsubsection{Collapse mitigants}
The initial version of DAC included codebook collapse mitigants, which were compatible with the initial VQ-VAE training trick: the straight through operator trick. These collapse mitigants are the normalization of codebooks and latents before quantization, and the projection of latents on a smaller dimensional space before quantization (the representation is projected back onto the initial space after quantization). The latter mitigant is performed thanks to convolutional layers placed at the input and the ouptut of each quantizer (see Fig. \ref{RandRVQ}). \\
We discuss in our experiments the impact of removing these mitigants for our model and the baseline. \\

\subsubsection{Model variants}
We propose five variations of our model to explore the impact of each parameter. The different parameters values characterizing each variant (from RandRVQ1 to RandRVQ5) are displayed in the upper part of Table \ref{metrics}. 
Through these experiments, a variation in the ratio of the size of the big codebook and the random sampling was introduced. We also explored the impact of the collapse mitigants, and the number of random quantizers.

\section{Results}
\label{sec-results}

\begin{table}[!t]
\renewcommand{\arraystretch}{1.3}
\caption{Reconstruction without (left) and with (right) randomized quantizers}
\label{part-reconstruction}
\centering
\begin{tabular}{ |c||c|c|} 
\hline
&Baseline 5q& RandRVQ1 \\

\hline

mel loss ($\downarrow$) &0.81& \textbf{0.72}  \\ 

stft loss ($\downarrow$) &2.10& \textbf{2.00}  \\ 

waveform loss ($\downarrow$) &0.05& \textbf{0.04}  \\

SI-SDR ($\uparrow$) &6.53& \textbf{8.25} \\ 

ViSQOL ($\uparrow$) &3.87& \textbf{3.94}  \\

\hline
\end{tabular}
\vspace{-2pt}
\end{table}

\begin{table*}[!t]
\renewcommand{\arraystretch}{1.3}
\caption{Objective evaluation on MUSDB18 dataset (averaged over 5 random draws)}
\label{metrics}
\centering
\begin{tabular}{ |c||c|c||c|c|c|c|c| } 
\hline
&Baseline& Collapse & RandRVQ1 & RandRVQ2 & RandRVQ3 & RandRVQ4 & RandRVQ5 \\

\hline

$N_{big}$ & $\emptyset$ & $\emptyset$ & 8192 & 8192 & 8192 & 16384 & 16384 \\

sample size &$\emptyset$& $\emptyset$ & 1024 & 1024 & 256 & 512 & 512  \\

Collapse miti. &$\surd$& $\emptyset$ & $\surd$ & $\emptyset$ & $\emptyset$ & $\surd$ & $\surd$  \\

\# rand. quantizers &0& 0 & 4 & 4 & 4 & 4 & 6  \\

\hline

mel loss ($\downarrow$) &0.71& 0.86 & \textbf{0.72$\pm$0.001} & 0.78$\pm$0.001 & 0.79$\pm$0.001 & 0.72$\pm$0.001 & 0.75$\pm$0.002  \\ 

stft loss ($\downarrow$) &1.99& 2.16 & \textbf{2.00$\pm$0.002} & 2.07$\pm$0.001 & 2.08$\pm$0.001 & 2.00$\pm$0.002 & 2.03$\pm$0.003 \\ 

waveform loss ($\downarrow$) &0.041& 0.057 & \textbf{0.042$\pm$0.000} & 0.048$\pm$0.000 & 0.048$\pm$0.000 & 0.043$\pm$0.0001 & 0.044$\pm$0.0001 \\

SI-SDR ($\uparrow$) &8.40& 5.26 & \textbf{8.25$\pm$0.02} & 6.99$\pm$0.01 & 7.00$\pm$0.01 & 8.14$\pm$0.02 & 7.76$\pm$0.03  \\ 

ViSQOL ($\uparrow$) &3.92& 3.85 & 
\textbf{3.94$\pm$0.007} & 3.87$\pm$0.006 & 3.88$\pm$0.004 & 3.93$\pm$0.008 & 3.92$\pm$0.01  \\

\hline

training time (days) &1.55& 1.583 & 1.459 & 1.465 & \textbf{1.442} & 1.462 & 1.51 \\

\hline
\end{tabular}
\end{table*}

\begin{table*}[!t]
\renewcommand{\arraystretch}{1.3}
\caption{Codebook perplexities (PP) on MUSDB18 Dataset (\textit{in parentheses: ratio to the maximum, trained codebooks are of size 1024})}
\label{perplexities}
\centering
\begin{tabular}{ |c||c|c||c|c|c|c|c| } 
    \hline
    
    & Baseline & Collapse & RandRVQ1 & RandRVQ2 & RandRVQ3 & RandRVQ4 & RandRVQ5 \\
    
    \hline

    \hline
    
    PP - cb 1 & 545 (0.53) & 25 (0.02) & 569 (0.55) & 536 (0.52) & 534 (0.52) & 545 (0.53) & 507 (0.49) \\
   
    PP - cb 2 & 834 (0.82) & 42 (0.04) & 810 (0.79) & 810 (0.79) & 817 (0.79) & 807 (0.78) & 828 (0.80) \\
    
    PP - cb 3 & 889 (0.86) & 67 (0.06) & 914 (0.89) & 873 (0.85) & 905 (0.88) & 896 (0.87) & 906 (0.88) \\
    
    PP - cb 4 & 928 (0.90) & 83 (0.08) & 913 (0.89) & 905 (0.88) & 924 (0.90) & 919 (0.89) & \textit{15565 (0.95)} \\
    
    PP - cb 5 & 935 (0.93) & 113 (0.11) & 947 (0.92) & 950 (0.92) & 927 (0.90) & 934 (0.91) & \textit{15586 (0.95)} \\
    
    PP - cb 6 & 946 (0.92) & 140 (0.13) & \textit{7744 (0.94)} & \textit{574 (0.07)} & \textit{3703 (0.45)} & \textit{15601 (0.95)} & \textit{15784 (0.96)} \\  
    
    PP - cb 7 & 957 (0.93) & 137 (0.13) & \textit{7720 (0.94)} & \textit{594 (0.07)} & \textit{3850 (0.46)} & \textit{15659 (0.95)} & \textit{15723 (0.95)} \\
    
    PP - cb 8 & 959 (0.93) & 166 (0.16) & \textit{7730 (0.94)} & \textit{616 (0.07)} & \textit{3997 (0.48)} & \textit{15622 (0.95)} & \textit{15923 (0.97)} \\
    
    PP - cb 9 & 966 (0.94) & 145 (0.14) & \textit{7776 (0.94)} & \textit{633 (0.07)} & \textit{4131 (0.50)} & \textit{15730 (0.96)} & \textit{15987 (0.97)} \\
    
    \hline
    
    PP - Big cb & $\emptyset$ & $\emptyset$ & \textit{7981 (0.97)} & \textit{606 (0.07)} & \textit{3937 (0.48)} & \textit{16090 (0.98)} & \textit{16230 (0.99)} \\
    
    \hline

    $N_{big}$ & $\emptyset$ & $\emptyset$ & \textit{8192} & \textit{8192} & \textit{8192} & \textit{16384} & \textit{16384} \\

    s & $\emptyset$ & $\emptyset$ & \textit{1024} & \textit{1024} & \textit{256} & \textit{512} & \textit{512} \\

    \hline
    
\end{tabular}
\end{table*}
The results\footnote{Some examples are displayed at: \url{https://randrvq.github.io/}} displayed in Table \ref{part-reconstruction} show the reconstruction metrics obtained with a partial Baseline (5 trained quantizers) and a total RandRVQ1 (5 trained and 4 randomized quantizers). It clearly shows that even using random codebooks without training, RandRVQ1 can still improve the reconstruction quality compared to a partial Baseline, underlining that deeper quantizers do encode useful information and that it is well captured by our random scheme. 

As shown in Table \ref{metrics}, the results obtained with our RandRVQ models are, overall, slightly below those of the Baseline. Yet, except for the SI-SDR metrics of RandRVQ2 and RandRVQ3, the figures are comparable, and even slightly better for VISQOL in RandRVQ1, RandRVQ4 and RandRVQ5.
These observations, even if they do not constitute a breakthrough in terms of quality of reconstruction, are positive.
Indeed, it indicates that in a setting where we forced the quantization to be untrained, and where the reconstruction is highly dependant on randomness, the obtained quality of the encoding we get from quantization at the same bitrate is globally comparable.

On the complexity side, as expected, our approaches demonstrate a slight advantage with a gain of a couple hours in training compared to the baseline. Nevertheless, this gain remains small. 

The codebook perplexities of the different models are given in Table \ref{perplexities}.
As a control experiment, we verify the mitigating effects of the collapse mitigants introduced in DAC \cite{kumar2024high}, as the collapse experiment shows a clear codebook collapse pattern in the perplexities, compared to the Baseline.

Looking at our experiments, we also notice that, overall, the behaviour of the learned first codebooks is similar to the Baseline, meaning that the usage of randomized quantizers does not influence these. Similarly, the perplexities measured for randomized codebooks which are featured with the collapse mitigants of \cite{kumar2024high}, indicate a nearly optimal usage of codebooks. This was predictable, as we are still profiting from the effect of the collapse mitigant, and as the quantization is subject to a random sampling which forces the exploration of the whole codebook.

The results from RandRVQ2 and RandRVQ3 bear more interesting information, as we can clearly see that in the case of no collapse mitigants, and poor parameters selection (RandRVQ2), the randomized quantizers fall back into codebook collapse in spite of the random sampling. This can be (at least partially) solved, by making a better choice of parameters, as we can see that RandRVQ3 has much better values of perplexities over its randomized quantizers (though not perfect, which indicates a possibility to probably go further). 

The difference between those two experiments comes from the sampling size that is applied for each random quantizer: the smaller the sampling, the smaller the choice of tokens for the quantizer, which forces exploration of the codebook, but limits the precision of the quantization. Then, the randomization of quantizers, associated with a correct choice of parameters, can be used as a collapse mitigant. 
Overall, these results are promising and indicate that the randomization of codebooks is potentially relevant for quantization since it can lead to good reconstruction metrics.

\section{Discussion}
\label{sec-discussion}

Although it is shown in this preliminary work that the randomization of quantizers bears promising prospects, some aspects must be further explored to better substantiate the  potential of our approach. 
\subsubsection{Fixed codebooks at inference} The "surprising" part of our method is that during inference, the random quantizers are still subject to random sampling. Such randomness could have great implications in terms of generalization, and it would be interesting to study the potential gain in exploiting several random draws to find an optimal quantization. 
Further explorations could be dedicated to finding a way to fix the sampling of the big codebook at inference. 

\subsubsection{Generalizability}
Preliminary experiments evaluating the generalization capabilities of our approach on unseen data (environmental audio data from ESC50) have shown a slightly better robustness than the baseline (although statistically insignificant). Future work is needed to fully explore the robustness potential of our approach.

\subsubsection{Training the big codebook}
Allowing the big codebook to be trained could further improve the reconstruction quality of our model, as the big codebook would still be of a much bigger size than traditional codebooks, and as it would converge through the training to a version of itself that would be adapted to the input data. It is also expected that codebook training may be necessary to allow an extension of the random process to the first codebooks which are apparently capturing more structured information departing clearly from Gaussian noise.

\vspace{-2pt}

\section{Conclusion}
\label{sec-conclusion}

Overall, though still in progress, this exploration has clearly shown that this novel concept of randomization of quantizers, in a context of quantized feature extraction, is very promising. 
As discussed above, there are several interesting directions that deserve to be pursued to better characterize and substantiate the potential of using random dictionaries including the design of optimal fixed dictionaries at inference, on the generalisation properties and on the extension of the random process to all codebooks of the Residual Vector quantization scheme.

\ifCLASSOPTIONcaptionsoff
  \newpage
\fi

\bibliographystyle{IEEEtran}
\bibliography{bibtex/abrv,bibtex/bib}

\end{document}